\documentclass[reqno,12pt]{amsart}

\usepackage{amssymb, amsmath}
\usepackage{fullpage}
\usepackage{graphics} 
\usepackage{epsfig} 

\def\const{\text{const.}}
\def\smallint{{\textstyle\int}}
\def\Hop{{\mathcal H}}
\def\X{{\mathrm X}}
\def\pr{{\mathrm pr}}

\newtheorem{thm}{Theorem}[section]

\title{Conservation laws and symmetries of\\ time-dependent generalized KdV equations}


\email{sanco@brocku.ca,}
\email{maria.rosa@uca.es}
\email{marialuz.gandarias@uca.es}

\thanks{$^*$ Corresponding author}

\begin{document}
\maketitle

\centerline{\scshape S. Anco}
\medskip
{\footnotesize
\centerline{Department of Mathematics and Statistics}
\centerline{Brock University}
\centerline{St. Catharines, Canada}
} 

\medskip

\centerline{\scshape M. Rosa, M.L. Gandarias$^*$}
\medskip
{\footnotesize
\centerline{Departamento de Matem\'aticas, Universidad  de C\'adiz}
\centerline{Pol\'igono del R\'io San Pedro s/n 11510 Puerto Real}
\centerline{C\'adiz, Spain}
}

\bigskip


\begin{abstract}
A complete classification of low-order conservation laws is obtained for 
time-dependent generalized Korteweg-de Vries equations. 
Through the Hamiltonian structure of these equations, 
a corresponding classification of Hamiltonian symmetries is derived. 
The physical meaning of the conservation laws and the symmetries is discussed. 
\end{abstract}

\section{Introduction}

This paper is devoted to a study of conservation laws and symmetries 
for a class of time-dependent generalized Korteweg-de Vries equations 
\begin{equation}\label{gkdv}
u_t +f(t,u)u_x + u_{xxx} =0
\end{equation}
with 
\begin{equation}\label{conds}
f|_{u=0}=0, 
\quad
f_{u}\neq 0 . 
\end{equation}
This class is preserved under the equivalence transformations
\begin{equation}\label{equivgroup}
t\rightarrow \tilde t = t + t_0, 
\quad
u\rightarrow \tilde u = u+u_0,
\quad
t_0, u_0 = \const . 
\end{equation}

Many interesting evolution equations are included in this class \eqref{gkdv}:
the KdV equation ($f=u$) models the dynamics of shallow water waves; 
the modified KdV equation $(f=u^2$) is a model for 
acoustic waves in anharmonic lattices \cite{mkdv1}
and Alfven waves in collision-free plasmas \cite{mkdv2}; 
a combined KdV-mKdV equation ($f=a u+ b u^2$, with $a$ and $b$ being arbitrary nonzero constants)
arises in plasma physics and solid-state physics, 
modelling wave propagation in nonlinear lattices \cite{kdv-mkdv1}
and thermal pulses in solids \cite{kdv-mkdv2,kdv-mkdv3}.
KdV-type equations having time-dependent coefficients 
arise in several applications \cite{timedep-kdv1,timedep-kdv2,timedep-kdv3,timedep-kdv4}
and can be mapped into this class \eqref{gkdv} by a point transformation  \cite{AncGan}.

All of these equations \eqref{gkdv} have a Hamiltonian structure, 
on any fixed spatial domain $\Omega\subseteq\mathbb{R}$, 
as given by 
\begin{equation}\label{hamilstruc}
u_t = \Hop(\delta H/\delta u)
\end{equation}
with the local Hamiltonian functional 
\begin{equation}\label{H}
H = \int_{\Omega} ( \tfrac{1}{2} u_x{}^2 - F(t,u) )\,dx, 
\quad
F(t,u) = u{\textstyle\int} f(t,u)\, du - {\textstyle\int} uf(t,u)\, du, 
\end{equation}
where the Hamiltonian operator \cite{Olv} is a total $x$-derivative 
\begin{equation}\label{Hop}
\Hop = D_x .
\end{equation}
The Hamiltonian $H$ will be a conserved integral when and only when 
the nonlinearity satisfies $f_t=0$,
which corresponds to a generalized KdV equation
$u_t + f(u)u_x + u_{xxx}=0$. 

Previous work on special families of equations in the class \eqref{gkdv} 
can be found in Refs.\cite{AncBlu02a,JohKhaBis,PopSer,MouKha}, 

In section~\ref{conslaws}, 
all low-order conservation laws of this class of generalized KdV equations \eqref{gkdv}--\eqref{conds}  
will be classified. 
In section~\ref{symms},
a corresponding classification of Hamiltonian symmetries will be derived
by using the Hamiltonian structure \eqref{hamilstruc} of the generalized KdV equations. 
The physical meaning of the symmetries and the conservation laws is discussed. 
Finally, some concluding remarks will be made in section~\ref{remarks}.

\section{Conservation laws}
\label{conslaws}

Conservation laws are also of basic importance in the study of evolution equations
because they provide physical, conserved quantities for all solutions $u(x,t)$,
and they can be used to check the accuracy of numerical solution methods
\cite{Olv,BluCheAnc}. 

A local conservation law for a time-dependent generalized KdV equation \eqref{gkdv} 
is a continuity equation 
\begin{equation}\label{conslaw}
D_t T+D_x X=0
\end{equation}
holding for all solutions $u(x,t)$ of equation \eqref{gkdv}, 
where the conserved density $T$ and the spatial flux $X$ are
functions of $t$, $x$, $u$, and $x$-derivatives of $u$. 
If $T=D_x\Theta$ and $X=-D_t\Theta$ hold for all solutions, 
then the continuity equation \eqref{conslaw} becomes an identity. 
Conservation laws of this form are called locally trivial,
and two conservation laws are considered to be locally equivalent if they differ
by a locally trivial conservation law. 
The global form of a non-trivial conservation law is given by 
\begin{equation}\label{globalconslaw}
\frac{d}{dt}\int_{\Omega} T\, dx = -X\Big|_{\partial\Omega}
\end{equation}
where $\Omega\subseteq\mathbb{R}$ is any fixed spatial domain. 

Every local conservation law can be expressed in an equivalent, characteristic form 
(analogous to the characteristic form for symmetries) \cite{Olv}
which is given by a divergence identity
\begin{equation}\label{chareqn}
D_t \tilde T+D_x \tilde X= (u_t +f(t,u)u_x +u_{xxx})Q
\end{equation}
holding off of the set of solutions of the evolution equation \eqref{gkdv}, 
where $\tilde T= T+D_x\Theta$ and $\tilde X= X-D_t\Theta$ 
are a conserved density and spatial flux that are locally equivalent to $T$ and $X$, 
and where \cite{AncBlu02b}
\begin{equation}\label{QTrel}
Q = E_u(\tilde T) 
\end{equation} 
is a function of $t$, $x$, $u$, and $x$-derivatives of $u$. 
This function is a called a multiplier \cite{Olv,AncBlu97,BluCheAnc}. 
Here $E_u$ denotes the Euler operator with respect to $u$ \cite{Olv}. 

For any evolution equation, 
there is a one-to-one correspondence between non-zero multipliers 
and non-trivial conservation laws up to local equivalence \cite{Olv,AncBlu02b},
and the conservation laws of basic physical interest arise from 
multipliers of low order \cite{Anc16a}. 
These multipliers for the evolution equation \eqref{gkdv} 
take the form 
\begin{equation}\label{low-order-Q}
Q(t,x,u,u_x,u_{xx})
\end{equation}
which correspond to conserved densities of the form 
\begin{equation}\label{low-order-T}
T(t,x,u,u_x) 
\end{equation}
modulo a trivial conserved density. 

A function \eqref{low-order-Q} will be a multiplier iff 
$E_u((u_t +f(t,u)u_x +u_{xxx})Q)=0$ holds identically,
since the kernel of the Euler operator consists of total divergences \cite{Olv,BluCheAnc}.
This condition splits with respects to any $x$-derivatives of $u$ that do not appear in $Q$.
The resulting overdetermined system consists of 
\begin{equation}\label{adjsymm-deteqn}
0=D_tQ + f D_xQ +D_x^3 Q
\end{equation}
and 
\begin{align}\label{helmholtz-deteqn}
Q_u = E_u(Q),
\
Q_{u_x} = -E_u^{(1)}(Q),
\
Q_{u_{xx}} = E_u^{(2)}(Q)
\end{align}
holding for all solutions $u(x,t)$ of the evolution equation \eqref{gkdv}. 
The first equation \eqref{adjsymm-deteqn} turns out to be 
the adjoint of the determining equation for symmetries 
(cf.\ \eqref{symm-deteqn}). 
The remaining equations \eqref{helmholtz-deteqn} constitute 
the Helmholtz equations \cite{AncBlu02b,Anc16a}
which are necessary and sufficient for $Q$ to have the variational form \eqref{QTrel}. 
Here $E_u^{(1)}$, and so on, denote the higher Euler operators \cite{Olv,Anc16a}. 

It is straightforward to set up and solve this determining system \eqref{adjsymm-deteqn}--\eqref{helmholtz-deteqn} 
subject to the classification conditions \eqref{conds}. 
The computation is simplest when we separate it into two main cases:
$f_{tu}=0$, and $f_{tu}\neq 0$. 
We merge the resulting subcases 
after first having solved the determining system in each of these two cases
and then having used the equivalence transformations \eqref{equivgroup}. 
(For solving the determining system, we use the Maple package ``rifsimp''.)

The multipliers \eqref{low-order-Q} for general $f(t,u)$ 
are linear combinations of 
\begin{align}
&
Q_1 =1;
\label{gencase1}
\\
&
Q_2 =u . 
\label{gencase2}
\end{align}
All special forms of $f(t,u)$ for which additional multipliers \eqref{low-order-Q} 
are admitted consist of:
\begin{subequations}
\begin{align}
&
f(t,u) =a(u), 
\quad
a(u) \text{ arbitrary} , 
\label{case1}
\\
& 
Q_3 = {-}u_{xx}  -\smallint a(u)\,du ;
\end{align}
\end{subequations}
\begin{subequations}
\begin{align}
&
f(t,u) = t^{-2/3} a(t^{1/3}u),
\quad
a(v) \text{ arbitrary} ,
\label{case2}
\\
& 
Q_4 = {-}t u_{xx} +\tfrac{1}{3}xu -t^{1/3} \smallint a(t^{1/3}u)\,du ;
\end{align}
\end{subequations}
\begin{subequations}
\begin{align}
& 
f(t,u) =a(t) u, 
\quad
a(t) \text{ arbitrary} ,
\label{case3}
\\
&
Q_5 =\big(\smallint a(t)\,dt\big) u -x  ;
\end{align}
\end{subequations}
\begin{subequations}
\begin{align}
&
f(u) =a(t) u, 
\quad
a(t) \text{  satisfies }
a^2 a''' -13 a'' a' a +24 a'{}^3 =0 , 
\label{case4}
\\
&
\begin{aligned}
Q_6 = & 
{-}2a(t)^{-3} u_{xx} -a(t)^{-2} u^2 
- 2x a'(t)a(t)^{-4}u 
\\&\qquad
{-}x^2(4 a'(t)^2 -a(t) a''(t))a(t)^{-6} ;
\end{aligned}
\end{align}
\end{subequations}
\begin{subequations}
\begin{align}
&
f(u) =a t^{-1/3} u + b u + c u^2, 
\quad
a, b, c \text{ constant} ,
\label{case5}
\\
&
\begin{aligned}
Q_7 = & 
{-}b t u_{xx} -\tfrac{1}{6} t( 2 c^2 u^3 +3cb u^2 + b^2 u ) 
\\&\qquad
{-}\tfrac{1}{4} a t^{2/3} (2c u^2+b u)  +\tfrac{1}{6} x (2c u +b) ;
\end{aligned}
\end{align}
\end{subequations}

Note, in the case \eqref{case4}, 
the third-order ODE possesses two first integrals 
$-2a^p a'' + (p+13)a^{p-1} a'{}^2 =c=\const$ for $p=-5$ and $p=-7$. 
This yields a reduction to a first-order separable ODE 
\begin{equation}
a'=a^3 \big(c_1 + c_2 a^2 \big)^{1/2} , 
\quad
c_1, c_2 \text{ constant}
\label{case4'}
\end{equation}
which has the quadrature 
\begin{equation}
\begin{aligned}
& \frac{2c_1^{1/2} + \big(c_1 + c_2 a^2 \big)^{1/2}}{a} \exp\Big(-\frac{c_1^{1/2}\big(c_1 + c_2 a^2 \big)^{1/2}}{c_2a^2}\Big) 
\\&
= \exp\Big( \frac{2c_1^{3/2}}{c_2} (t+c_3) \Big), 
\quad
c_1, c_2, c_3 \text{ constant}
\end{aligned}
\label{case4''}
\end{equation}

For each multiplier admitted by a time-dependent generalized KdV equation \eqref{gkdv},
a corresponding conserved density and flux can be derived (up to local equivalence) 
by integration of the divergence identity \eqref{chareqn} 
\cite{BluCheAnc,Anc16a}. 
We obtain the following results. 

\begin{thm}\label{classify-conslaws}
(i) All conservation laws given by low-order conserved densities \eqref{low-order-T}
admitted by the class of time-dependent generalized KdV equations \eqref{gkdv} 
for arbitrary $f(t,u)$ (satisfying conditions \eqref{conds})
are linear combinations of:
\begin{flalign}
&\label{conslaw1}
T_1= u ,
\quad
X_1= u_{xx} +\smallint f(t,u)\,du ;
&
\\
&\label{conslaw2}
T_2= \tfrac{1}{2}u^2 ,
\quad
X_2= uu_{xx}-\tfrac{1}{2}u_x^2 +\smallint uf(t,u)\,du .
&
\end{flalign}
(ii) The class of time-dependent generalized KdV equations \eqref{gkdv} 
admits additional conservation laws given by low-order conserved densities \eqref{low-order-T}
only for $f(t,u)$ of the form \eqref{case1}, \eqref{case2}, \eqref{case3}, \eqref{case4}
(satisfying conditions \eqref{conds}). 
The admitted conservation laws in each case are given by:
\begin{subequations}
\label{conslaw3}
\begin{flalign}
&
T_3 = \tfrac{1}{2}u_x{}^2  -\smallint A(u)\,du , 
&
\\
&
X_3 = {-}\tfrac{1}{2}u_{xx}{}^2 -A(u) u_{xx} -u_tu_x -\tfrac{1}{2} A(u)^2 , 
&
\\
& A(u) = \smallint a(u)\,du ;
\end{flalign}
\end{subequations}
\begin{subequations}
\label{conslaw4}
\begin{flalign}
&
\begin{aligned}
T_4 & = \tfrac{1}{2} t u_x{}^2 +\tfrac{1}{6} x u^2  -\smallint A(t^{1/3}u)\,du ,
\end{aligned}
&
\\
&
\begin{aligned}
X_4 & = -\tfrac{1}{2} t u_{xx}{}^2 - A(t^{1/3}u) u_{xx} 
+ \tfrac{1}{6} x( 2uu_{xx} - u_x{}^2 )
\\&\qquad 
+ \tfrac{1}{3} x t^{-1} \big( uA(t^{1/3}u) -\smallint A(t^{1/3}u)\,du \big)
-\tfrac{1}{2} t^{-1} A(t^{1/3}u)^2 , 
\end{aligned}
&
\\
& A(v) = \smallint a(v)\,dv ;
\end{flalign}
\end{subequations}
\begin{subequations}
\label{conslaw5}
\begin{flalign}
& 
T_5 = \tfrac{1}{2} A(t)  u^2 - x u , 
&
\\
&
X_5 = {-}\tfrac{1}{2} x( 2u_{xx} +a(t) u^2 )  +u_x  
+\tfrac{1}{2} A(t) ( 2u u_{xx} -u_x{}^2 )
- \tfrac{1}{3} a(t) A(t) u^3 ,
&
\\
&
A(t) =\smallint a(t)\, dt ;
&
\end{flalign}
\end{subequations}
\begin{subequations}
\label{conslaw6}
\begin{flalign}
& 
\begin{aligned}
T_6 & = a(t)^{-3} u_x{}^2 -\tfrac{1}{3} a(t)^{-2} u^3
-c_1 x^2 u - \big( c_2 + c_1 a(t)^{-2} \big)^{1/2} x u^2 , 
\end{aligned}
&
\\
& 
\begin{aligned}
X_6 & = {-}a(t)^{-3} (u_{xx}{}^2+2 u_t u_x) 
- a(t)^{-2} u^2 u_{xx} 
+2\big( c_2 + c_1 a(t)^{-2} \big)^{1/2} u u_x
\\&\qquad
{-}\tfrac{1}{4} a(t)^{-1} u^4 +2c_1 (u-xu_x)
-\tfrac{1}{2}c_1 x^2( 2u_{xx} + a(t) u^2)
\\&\qquad
{-} x \big( c_2 + c_1 a(t)^{-2} \big)^{1/2} ( 2u u_{xx} - u_x{}^2 )
-\tfrac{2}{3} x \big( c_1 + c_2 a(t)^{2} \big)^{1/2} u^3 ;
\end{aligned}
&
\end{flalign}
\end{subequations}
\begin{subequations}
\label{conslaw7}
\begin{flalign}
& 
\begin{aligned}
T_7 & = \tfrac{1}{2} c t u_x{}^2 -\tfrac{1}{12} t ( c u^2 + b u )^2
+\tfrac{1}{6} x( c u^2 + b u ) - \tfrac{1}{2} a t^{2/3} ( \tfrac{1}{3}c u^3 + \tfrac{1}{4} b u^2 ) , 
\end{aligned}
&
\\
& 
\begin{aligned}
X_7 & = {-}\tfrac{1}{2} c t (u_{xx}{}^2 +2 u_tu_x) 
-\tfrac{1}{6} t ( 2c^2 u^3 +3 bc u^2 + b^2 u )u_{xx}
+\tfrac{1}{12} t b^2 u_x{}^2 
\\&\qquad
-\tfrac{1}{18} t ( c u^2 +b u )^3 +\tfrac{1}{12} x ( c u^2 + b u)^2  
+\tfrac{1}{6} x ( ( 2c u + b) u_{xx} -b u_x{}^2 ) 
\\&\qquad
+\tfrac{1}{3}a x t^{-1/3} (\tfrac{1}{3} c u^3 + \tfrac{1}{4} b u^2) 
-\tfrac{1}{12} a^2 t^{1/3} (\tfrac{3}{2}c u^4 + b u^3) 
\\&\qquad
-\tfrac{1}{4} a t^{2/3} ( (2cu^2 + b u)u_{xx} - \tfrac{1}{2}b u_x{}^2 
+ 2c^2 u^5 + \tfrac{5}{4} bc u^4 + \tfrac{5}{3} b^2 u^3 )
\\&\qquad
-\tfrac{1}{6} ( 2c u + b)  u_x . 
\end{aligned}
&
\end{flalign}
\end{subequations}
\end{thm}

Note that $u_t$ can be eliminated in the spatial flux expressions 
by use of the evolution equation \eqref{gkdv}. 

The physical meaning of these conservation laws \eqref{conslaw1}--\eqref{conslaw7} 
can be seen by considering their global form \eqref{globalconslaw}. 

For general $f(t,u)$, 
the two admitted conservation laws \eqref{conslaw1} and \eqref{conslaw2}
yield the conserved integrals 
\begin{align}
&\label{globalconslaw1}
C_1 = \int_{\Omega} u\, dx , 
\\
&\label{globalconslaw2}
C_2 = \int_{\Omega} \tfrac{1}{2}u^2\, dx . 
\end{align}
These represent the total mass and the $L^2$-norm for solutions $u(x,t)$. 

In the time-independent case \eqref{case1}, 
where $f(t,u) =a(u)=A'(u)$, 
the conservation law \eqref{conslaw3} yields the conserved integral 
\begin{equation}\label{globalconslaw3}
C_3 = \int_{\Omega} \big( \tfrac{1}{2}u_x{}^2 -\smallint A(u)\,du \big)\,dx
\end{equation}
which represents the Hamiltonian or the total energy for solutions $u(x,t)$. 

In the time-dependent nonlinear case \eqref{case2}, 
where $f(t,u) = t^{-2/3} a(t^{1/3}u) = t^{-2/3} A'(t^{1/3}u)$, 
the conservation law \eqref{conslaw4} yields the conserved integral 
\begin{equation}\label{globalconslaw4}
C_4 = \int_{\Omega} \big( \tfrac{1}{2} t u_x{}^2 +\tfrac{1}{6} x u^2 - \smallint A(t^{1/3}u)\,du \big)\,dx
\end{equation}
which represents a dilational energy for solutions $u(x,t)$. 

In the time-dependent linear cases \eqref{case3} and \eqref{case4}, 
the two conservation laws \eqref{conslaw5} and \eqref{conslaw6} 
respectively yield the conserved integrals
\begin{equation}\label{globalconslaw5}
C_5 = \int_{\Omega} \big( \tfrac{1}{2} A(t)  u^2 -x u \big)\,dx
\end{equation}
where $f(t,u) =a(t) u =A'(t)u$,
and 
\begin{equation}\label{globalconslaw6}
C_6 = \int_{\Omega} \big( a(t)^{-3} u_x{}^2 -\tfrac{1}{3} a(t)^{-2} u^3 
-\big( c_2 + c_1 a(t)^{-2} \big)^{1/2} x u^2 -c_1 x^2 u \big)\,dx
\end{equation}
where $f(t,u) =a(t) u$ with $a(t)$ given by expression \eqref{case4''}. 
Since $f(t,u)$ is linear in $u$ in these two cases, 
the evolution equation \eqref{gkdv} has the form
\begin{equation}\label{timedepKdV}
u_t + a(t)uu_x + u_{xxx}=0
\end{equation}
which is a KdV equation with a time-dependent coefficient, 
where $a(t) u$ physically represents an advective velocity.  
Then the first conserved integral \eqref{globalconslaw5} 
describes a generalized Galilean momentum,
and the second conserved integral \eqref{globalconslaw6} 
describes a generalized dilational energy. 
In particular, when $a=\const$, these conserved integrals reduce to 
the ordinary Galilean momentum 
$\smallint_{\Omega} \big( \tfrac{1}{2} a t u^2 -x u \big)\,dx$
and the ordinary energy 
$a^{-3} \smallint_{\Omega} \big( u_x{}^2 - a u^3 \big)\,dx$
for the KdV equation. 

In the quadratic case \eqref{case5}, 
where $f(u) =a t^{-1/3} u + b u + c u^2$, 
the conservation law \eqref{conslaw7} yields the conserved integral 
\begin{equation}\label{globalconslaw7}
C_7 = \int_{\Omega} \big( \tfrac{1}{2} c t u_x{}^2 -\tfrac{1}{12} t ( c u^2 + b u )^2
+\tfrac{1}{6} x( c u^2 + b u ) - \tfrac{1}{24} a t^{2/3} ( 4c u^3 + 3 b u^2 )\big)\,dx
\end{equation}
which represents a combined Galilean energy-momentum for solutions $u(x,t)$. 
In particular, when $a=b=0$, 
this conserved integral reduces to the Galilean energy 
$c \smallint_{\Omega} \big( \tfrac{1}{2} t u_x{}^2 -\tfrac{1}{12} c t u^4 +\tfrac{1}{6} x u^2 \big)\,dx$
for the mKdV equation,
while when $a=c=0$, 
the Galilean momentum for the KdV equation is obtained.

\section{Symmetries}
\label{symms}

Symmetries are a basic structure of evolution equations 
as they can be used to find invariant solutions
and yield transformations that map the set of solutions $u(x,t)$ into itself
\cite{Olv,BluCheAnc}.

An infinitesimal symmetry for a time-dependent generalized KdV equation \eqref{gkdv} 
is a generator 
\begin{equation}\label{generator}
\X =\xi \partial_x+\tau \partial_t+\eta \partial_u
\end{equation} 
whose prolongation leaves invariant the equation \eqref{gkdv},
where $\xi$, $\tau$, and $\eta$ are functions of $t$, $x$, $u$, and $x$-derivatives of $u$.
The symmetry is trivial if it leaves invariant every solution $u(x,t)$ of the equation \eqref{gkdv}. 
This occurs when (and only when) $\xi$, $\tau$, and $\eta$ satisfy the relation
\begin{equation}
\eta=u_x \xi + u_t \tau
\end{equation}
for all solutions $u(x,t)$.  
The corresponding generator \eqref{generator} of a trivial symmetry
is given by
\begin{equation}\label{trivsymm}
\X_{\rm triv}= \xi \partial_x + \tau \partial_t + (\xi u_x + \tau u_t)\partial_u
\end{equation}
which has the prolongation $\pr\X_{\rm triv}=\xi D_x + \tau D_t$

Any symmetry generator is equivalent \cite{Olv,BluCheAnc} to a generator 
\begin{equation}\label{symmchar}
\hat\X=\X-\X_{\rm triv} = P\partial_u, 
\quad
P=\eta -\xi u_x -\tau u_t
\end{equation}
under which $u$ is infinitesimally transformed while $x$ and $t$ are invariant,
due to the relation
\begin{equation}
\pr\X-\pr\hat\X= \xi D_x + \tau D_t . 
\end{equation}
This generator \eqref{symmchar} defines the \emph{characteristic form} 
for the infinitesimal symmetry.
Invariance of the evolution equation \eqref{gkdv} is 
then given by the condition \cite{Olv,Anc16a}
\begin{equation}\label{symm-deteqn}
0=D_tP +D_x(f P) +D_x^3 P
\end{equation}
holding for all solutions $u(x,t)$ of the equation \eqref{gkdv}. 

A symmetry will generate a point transformation on $(x,t,u)$ iff 
the coefficients $\xi$, $\tau$, and $\eta$ in its characteristic function $P$ 
depend only on $x,t,u$ \cite{Olv,BluCheAnc}, 
yielding a generator with the form 
\begin{equation}
\X =\xi(x,t,u)\partial_x+\tau(x,t,u)\partial_t+\eta(x,t,u)\partial_u . 
\end{equation} 

For any Hamiltonian evolution equation, 
there is a correspondence 
that produces a symmetry from each admitted conservation law. 
This correspondence is a Hamiltonian analog of Noether's theorem. 
It can be formulated for an evolution equation \eqref{gkdv}, 
with the Hamiltonian structure \eqref{Hop}, 
through the explicit relation \cite{Olv,MaZho} 
\begin{equation}\label{PQmap}
P = \Hop(\delta C/\delta u) = D_x Q
\end{equation} 
involving the characteristic function $P$ of the symmetry \eqref{symmchar}
and the multiplier $Q$ associated to the conserved integral 
$C = \smallint_{\Omega} T\, dx$
given by a conservation law \eqref{conslaw}. 
This correspondence is one way: every conservation law yields a symmetry. 
The converse holds iff the symmetry has the form \eqref{PQmap}, 
which requires that $E_u(P)=0$. 

The classification of low-order conservation laws stated in Theorem~\ref{classify-conslaws}
for the class of time-dependent generalized KdV equations \eqref{gkdv} 
yields the following symmetries 
produced from the conservation law multipliers. 

The multipliers \eqref{gencase1} and \eqref{gencase2} 
given by the two conservation laws \eqref{conslaw1} and \eqref{conslaw2}
admitted for general $f(t,u)$ 
produce the symmetry characteristic functions
\begin{align}
&
P_1 =0;
\\
&
P_2 =u_x . 
\label{symm1,2}
\end{align}
The additional multipliers \eqref{case1}--\eqref{case5}
given by the conservation laws \eqref{conslaw3}--\eqref{conslaw7}
which are admitted for forms of $f(t,u)$
respectively yield the symmetry characteristic functions
\begin{align}
&
P_3 = {-}u_{xxx}  -a(u)u_x ;
\label{symm3}
\\
& 
P_4 = {-}t u_{xxx} +\tfrac{1}{3}xu_x +\tfrac{1}{3}u  -t^{2/3} a(t^{1/3}u)u_x ;
\label{symm4}
\\
&
P_5 =\big(\smallint a(t)\,dt\big) u_x -1  ;
\label{symm5}
\\
& 
\begin{aligned}
P_6 = & 
{-}2a(t)^{-3} u_{xxx} -2a(t)^{-2} uu_x 
- 2x a'(t)a(t)^{-4}u_x 
\\&\qquad
- 2 a'(t)a(t)^{-4}u 
-2x(4 a'(t)^2 -a(t) a''(t))a(t)^{-6} ;
\end{aligned}
\label{symm6}
\\
&
\begin{aligned}
P_7 = & 
{-}b t u_{xxx} -t( c^2 u^2 +cb u + \tfrac{1}{6} b^2 ) u_x
\\&\qquad
{-}a t^{2/3} (c u+\tfrac{1}{4} b)u_x  +\tfrac{1}{3} c x u_x
+\tfrac{1}{6} (2c u +b) .
\end{aligned}
\label{symm7}
\end{align}

By evaluating these characteristic functions \eqref{symm1,2}--\eqref{symm7}
on solutions $u(x,t)$, 
we can eliminate all $u_{xxx}$ terms to obtain 
$P(t,x,u,u_x,u_t) =\eta(t,x,u) -\xi(t,x,u) u_x -\tau(t,x,u) u_t$
in each case. 
This leads to the following symmetry classification. 

\begin{thm}\label{classify-hamilsymms}
(i) The symmetries corresponding to the two low-order conserved integrals \eqref{globalconslaw1}--\eqref{globalconslaw2} 
admitted by the class of time-dependent generalized KdV equations \eqref{gkdv} 
for arbitrary $f(t,u)$ (satisfying conditions \eqref{conds})
are generated by:
\begin{equation}
\label{generator1,2}
\X_1 =0,
\quad
\X_2 = \partial_x . 
\end{equation}
(ii) The additional symmetries corresponding to the low-order conserved integrals \eqref{globalconslaw3}--\eqref{globalconslaw7} 
admitted for special forms of $f(t,u)$ (satisfying conditions \eqref{conds})
are generated in each case by:
\begin{subequations}
\label{generator3}
\begin{flalign}
&
\X_3 = \partial_t ,
&
\\
&
f(t,u) =a(u), 
\quad
a(u) \text{ arbitrary} ;
\label{symmcase1}
&
\end{flalign}
\end{subequations}
\begin{subequations}
\label{generator4}
\begin{flalign}
&
\X_4 = \tfrac{1}{3}x\partial_x + t\partial_t +\tfrac{1}{3}u \partial_u ,
&
\\
& 
f(t,u) = t^{-2/3} a(t^{1/3}u),
\quad
a(v) \text{ arbitrary} ;
\label{symmcase2}
&
\end{flalign}
\end{subequations}
\begin{subequations}
\label{generator5}
\begin{flalign}
& 
\X_5 =a(t)\partial_x -\partial_u  ,
&
\\
&
f(t,u) =a(t) u, 
\quad
a(t) \text{ arbitrary} ;
\label{symmcase3}
&
\end{flalign}
\end{subequations}
\begin{subequations}
\label{generator6}
\begin{flalign}
& 
\begin{aligned}
\X_6 & = - 2 a'(t)a(t)^{-4} x\partial_x  +2 a(t)^{-3}\partial_t 
\\&\qquad
+\big( 2x (a(t) a''(t)-4 a'(t)^2)a(t)^{-6}  -2a'(t)a(t)^{-4}\big)\partial_u ,
\end{aligned}
&
\\
&
f(u) =a(t) u, 
\quad
a(t) \text{  satisfies }
a^2 a''' -13 a'' a' a +24 a'{}^3 =0 ;
\label{symmcase4}
&
\end{flalign}
\end{subequations}
\begin{subequations}
\label{generator7}
\begin{flalign}
& 
\begin{aligned}
\X_7 = & (\tfrac{1}{3}c x - \tfrac{1}{6}b^2 t - \tfrac{1}{4} ab t^{2/3})\partial_x 
+ c t\partial_t +(\tfrac{1}{3}c u + \tfrac{1}{6}b)\partial_u ,
\end{aligned}
&
\\
&
f(u) =a t^{-1/3} u + b u + c u^2, 
\quad
a, b, c \text{ constant} .
\label{symmcase5}
&
\end{flalign}
\end{subequations}
\end{thm}
 
Note, from the quadrature \eqref{case4''} for the ODE for $a(t)$ in case \eqref{generator6}, 
we can express 
\begin{equation}
\X_6 = -2\big( c_2 + c_1 a(t)^{2} \big)^{1/2} x\partial_x + 2a(t)^{-3}\partial_t 
-2\big(c_1 x + \big( c_2 + c_1 a(t)^{2} \big)^{1/2}\big)\partial_u .
\end{equation}

All of the symmetries \eqref{generator1,2}--\eqref{generator7} are point symmetries. 
Their physical meaning will now be discussed. 

For general $f(t,u)$, 
the symmetry $\X_1$ obtained from the conservation law \eqref{conslaw1} is trivial. 
Consequently, the conservation law \eqref{conslaw1} represents 
a Casimir of the Hamiltonian structure \cite{Olv}. 
The other symmetry $\X_2$ is a space translation. 

In the time-independent case \eqref{symmcase1}, 
where the conserved integral \eqref{globalconslaw3} 
represents the Hamiltonian or the total energy for solutions $u(x,t)$, 
the symmetry $\X_3$ is a time translation. 

In the time-dependent nonlinear case \eqref{symmcase2}, 
where the conserved integral \eqref{globalconslaw4} 
represents a dilational energy for solutions $u(x,t)$, 
the symmetry $\X_4$ is a scaling. 

In the time-dependent linear cases \eqref{symmcase3} and \eqref{symmcase4}, 
where the two conserved integrals \eqref{globalconslaw5} and \eqref{globalconslaw6} 
respectively represent a generalized Galilean momentum
and a generalized dilational energy, 
the first symmetry $\X_5$ is a generalized Galilean boost
and the second symmetry $\X_6$ is a generalized dilation. 
Note the evolution equation \eqref{gkdv} in these cases has the form of 
a time-dependent KdV equation \eqref{timedepKdV}
in which $a(t) u$ physically represents an advective velocity.  
In particular, when $a=\const$, these two symmetries reduce to 
an ordinary Galilean boost and a time translation. 

In the quadratic case \eqref{symmcase5}, 
where the conserved integral \eqref{globalconslaw7} 
represents a combined Galilean energy-momentum for solutions $u(x,t)$,
the symmetry $\X_7$ is a scaling combined with a Galilean boost. 
In particular, when $a=b=0$, 
this symmetry reduces to the scaling symmetry for the mKdV equation,
while when $a=c=0$, 
the Galilean boost symmetry for the KdV equation is obtained.

\section{Concluding remarks}
\label{remarks}

The classifications of low-order conservation laws and associated Hamiltonian symmetries
obtained in this paper can be extended to a wider class of evolution equations 
\begin{equation}
u_t + f(t,u) u_x + b(t)u + c(t)u_{xxx} = s(t) 
\end{equation}
by use of a mapping 
\begin{equation}
x\rightarrow \tilde x = x - \zeta(t),
\quad
t\rightarrow \tilde t = \tau(t), 
\quad
u\rightarrow \tilde u = \lambda(t) u+\nu(t) 
\end{equation}
with $\tau'(t)\neq 0$ and $\lambda(t)\neq 0$. 
This will be carried out in a subsequent paper \cite{AncGan}.

\section*{Acknowledgments} 
M.R.\ and M.L.G.\ gratefully acknowledge the support of Junta de Andaluc\'ia group FQM-201.
S.C.A.\ is supported by an NSERC research grant.

\medskip
\medskip

\end{document}